\documentstyle[twocolumn,aps,psfig,epsfig] {revtex}
\begin{document}
\bibliographystyle{unsrt}

\title{Origins of Intermediate Velocity Particle Production
in Heavy Ion Reactions}

\author{L. Gingras$^1$,
A. Chernomoretz$^2$,
Y. Larochelle$^1$\thanks{Present address: Joint Institute for Heavy Ion Research,
Holifield Radioactive Ion Beam Facility, Oak Ridge, Tennessee 37831-6374.},
Z.Y. He$^1$\thanks{Present address: EXFO, 465 avenue Godin, Vanier, QC, Canada G1M 3G7.},
L. Beaulieu$^1$, G.C. Ball$^3$\thanks{Present address: TRIUMF, 4004 Wesbrook Mall, Vancouver, B.C., Canada, V6T 2A3},
F. Grenier$^1$, D. Horn$^3$,
R. Roy$^1$, M. Samri$^4$, C. St-Pierre$^1$, D. Th\'eriault$^1$ and S. Turbide$^1$}
\address{$^1$ Laboratoire de Physique Nucl\'eaire, D\'epartement de Physique, Universit\'e Laval, Qu\'ebec, QC, Canada G1K 7P4.}
\address{$^2$ Departamento de Fisica, Universidad de Buenos Aires, Buenos Aires, Argentina.}
\address{$^3$ AECL, Chalk River Laboratories, Chalk River, ON, Canada K0J 1J0.}
\address{$^4$ Laboratoire de Physique Nucl\'eaire et Applications, D\'epartement de Physique,
Universit\'e Ibn Tofail, K\'enitra, Morocco.}

\date{\today}
\maketitle

\begin{abstract}
Investigation of intermediate-velocity particle production is performed on
entrance channel mass asymmetric collisions of $^{58}$Ni+C and $^{58}$Ni+Au
at 34.5 MeV/nucleon. Distinctions between prompt pre-equilibrium ejections,
multiple neck ruptures and an alternative phenomenon of delayed aligned
asymmetric breakup is achieved using source reconstructed correlation
observables and time-based cluster recognition in molecular dynamics simulations.

\end{abstract}
PACS number(s): 25.70 -z, 25.70.Lm, 25.70.Mn, 25.70.Pq

During the last decade, reaction mechanisms studies of heavy ion collisions in
the Fermi energy domain have unraveled new and interesting phenomena that can
have major consequences on our understanding of both nuclear dynamics and
excited nuclear matter properties. It was first established that binary dissipative
collisions dominate the reaction cross section in this energy range for a large
domain of entrance channel masses and asymmetries \cite{lott92,leco94,pete95,laro95,beau96b}.
Early, it has been noticed that unlike the lower energy case, where
exclusively two primary fragments (or one fusion fragment) are formed \cite{schr84},
the intermediate energy heavy ion collisions produce exotic structures that
populate the intermediate-velocity (IV) zone between the two main collision partners
\cite{mont94,leco95,toke95,demp96,laro97b,luka97,lefo00,plag00,poggi01}.
This production of particles and fragments at mid-velocity is usually explained as
the onset of the participant-spectator model of relativistic energies \cite{west76}.
However, in the Fermi energy domain, the interplay between two-body interactions
and collective motions from the nuclear mean field modifies this simple picture
and can potentially give rise to peculiar mechanisms leading to such an important
observed IV particle production. Of those mechanisms, fast light particles
and coalescence clusters ejected by nucleon-nucleon collisions and multiple neck
ruptures are usually expected to be responsible for the main IV particle production.

In this paper, we report for the first time clear distinctions between these prompt
processes and an alternative phenomenon of delayed aligned asymmetric breakup
that populates the intermediate-velocity zone by
a deformation rupture of mainly the heavier of the colliding partners in
mass asymmetric collisions. These distinctions were observed experimentally with
intermediate-velocity particle correlation analysis of Ni+C and Ni+Au reactions
at 34.5 MeV/nucleon.

A beam of $^{58}$Ni accelerated at 34.5 MeV/nucleon by the coupled Tandem and SuperConducting
Cyclotron accelerators of AECL at Chalk River bombarded alternatively a
2.4 mg/cm$^{2}$ carbon target and a 2.7 mg/cm$^{2}$ gold target.
Charged particles issued from these reactions were detected in the CRL-Laval 4$\pi$ array
constituted by 144 detectors set in ten rings concentric to the beam axis and
covering polar angles between 3.3$^o$ and 140$^o$.
The first forward four rings (3.3$^o$ to 24$^o$) are each made of 16 plastic phoswich detectors
with energy detection thresholds of 7.5 (27.5) MeV/nucleon for element identification of Z=1 (28)
particles. Between 24$^o$ and 46$^o$, two rings of 16 CsI(Tl) crystals achieve isotopic
resolution for Z=1 and 2 ions and element identification for Z=3 and 4 ions
with energy thresholds ranging from 2 to 5 MeV/nucleon. The Miniball forms the last
four rings (46$^o$ to 87$^o$ and 93$^o$ to 140$^o$ ) and is constituted by PIN diode
backed CsI(Tl) crystal detectors set in
groups of 12 per ring.
See \cite{ging98a} and references therein for more information on detectors and energy calibration.
The main trigger for event recording was a charged particle multiplicity
of at least 3 particles.
The present work is restricted to events selected in the off-line analysis
by a total detected charge of at least 31 units.
This ensures good reconstruction of the
quasi-projectile and IV particles characteristics.

In order to evaluate properties of the IV material, it needs to be
isolated from its surrounding. This is done by reconstructing on
an event-by-event basis the quasi-projectile (QP) and quasi-target (QT) emitters using
a probabilistic reconstruction algorithm for the first nucleus and angular and velocity cuts for
emissions of the second one. The statistical quasi-projectile reconstruction
algorithm proceeds by two steps consisting in building probability tables
for the attribution to the QP of a final detected particle and an application
of these probabilities on an event-by-event basis. The main hypotheses on which
the algorithm relies on are: (1) the event heaviest particle is assumed to be the QP 
evaporation residue, (2) particles emitted by the QP are isotropically distributed around this
residue, and (3) the forward emission hemisphere of the QP is negligeably contaminated by 
emissions from other processes. Clearly, these hypotheses can only be verified in peripheral to
semi-central collisions where the QP and QT are well enough separated in velocity space.
Elements of the probability tables are referenced by classes of experimental
impact parameters ($b_{exp}$), residue ($Z_{res}$) and emitted particle ($Z_p$)
atomic numbers, as well as relative velocity between them ($V_{rel}$) and its absolute projection along
the residue velocity in the CM ($V_{rel}^{||}$).
The experimental impact parameter was determined using a combination of two observables
that are both closely related to the true impact parameter according to molecular dynamics
and deep inelastic simulations \cite{tass91}.
These observables are the total parallel momentum of all charged particles in the forward velocity hemisphere
of the center of mass (CM) reference frame ($\Pi_{||}^{cm}$) and the anisotropy ratio of light charged particles
($Z \leq 2$) in that same frame ($R_A$) \cite{stro83}.
\begin{equation}
b_{exp}=r_0 (A_P^{1/3} + A_T^{1/3}) \frac{1 + \frac{\Pi_{||}^{cm}}{P_P^{cm}} - c \frac{tg^{-1}(R_A)}{\pi/2}}{2},
\end{equation}
where $A_P$, $A_T$ and $P_P^{cm}$ are respectively the projectile mass number, target mass number and
projectile CM momentum. $r_0 = 1.2 fm$ and the constant $c$ was experimentally adjusted
to 1.2.
Following hypotheses
(2) and (3), the QP attribution probability for particles emitted in the forward
hemisphere is fixed at unity, whereas the probability for the backward-emitted
ones is determined by dividing the forward with the backward relative velocity spectra
gated by $b_{exp}$, $Z_{res}$, $Z_p$ and $V_{rel}^{||}$. Details and results of this method for QP properties can be
found in refs. \cite{ging98a,ging98b,he00}. Particles detected in the Miniball
were attributed to the QT emitter as well as particles having parallel velocity along the residue
direction of less than $-0.12 c$. All remaining final particles are
assigned to the IV component.

\begin{figure}
\centerline{\epsfig{figure=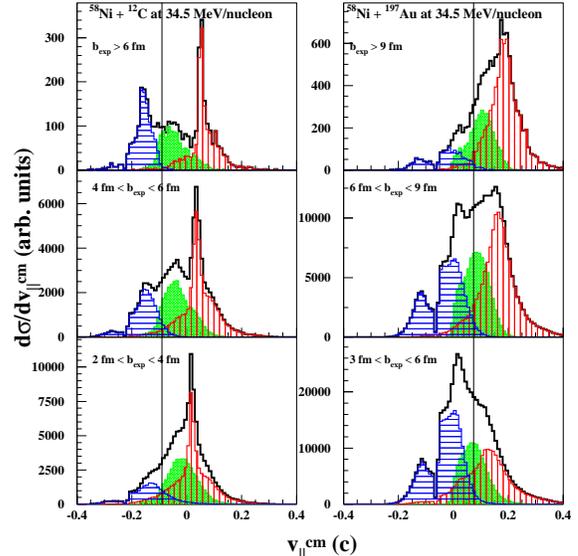,width=7.5cm}}
\caption{\label{fig_vpar} CM Parallel velocity distributions for charged
particles of $^{58}$Ni+C (left) and $^{58}$Ni+Au (right) reactions. The total
distributions (thick solid lines) are broken in QP (vertical hatches),
QT (horizontal hatches) and IV (shaded histogram) contributions.
The vertical line shows $v_{NN}$.}
\end{figure}

Figure \ref{fig_vpar} shows the evolution with experimental impact parameter of the
parallel velocity distribution in the CM reference frame
of the reaction charged products according to their attributed origin.
In both reactions, as the impact parameter decreases, the mean parallel velocity of
the QP products also decreases, reflecting the energy damping of the
projectile. Along with this obvious effect, particle production from the QT and IV
material, for $^{58}$Ni+Au reaction, becomes more important as the collision
gets more violent. In the case of $^{58}$Ni+C reaction, the $\alpha$ particle sub-structure
of the $^{12}$C target makes it easy to break at all detected impact parameters.
Particle production in the IV region appears also to be strong on
the whole $b_{exp}$ range. With the decrease of impact parameter, the average velocity
of the IV charged products in the $^{58}$Ni+C system increases
above $v_{NN}$, whereas for $^{58}$Ni+Au system, it stays close to it on the whole range, with
a small tendancy to decrease toward the QT velocity.

In order to get information on the time formation and characteristics of the IV
material produced in these mass asymmetric heavy ion reactions,
molecular dynamics simulations were performed. 
A semi-classical molecular dynamics model with a spherical two-body
interaction potential presenting realistic nucleon-nucleon cross sections
and including Coulomb interactions has been used to modelize semi-peripheral reactions
of the $^{58}$Ni+$^{12}$C and $^{58}$Ni+$^{197}$Au systems at $34.5$ MeV/nucleon.
Focusing on the nucleon-nucleon correlations that give rise to the 
fragment formation processes, a clusterization algorithm making use of
the `minimum spanning tree in energy space' (MSTE) criterion, has been used.
Right after the most violent stage of the collision, an MSTE cluster recognition 
step allowed us to sort the particles as belonging to the projectile-like 
fragment (PLF), the target-like fragment (TLF), or as prompt emitted particles (PEP)
not spatially correlated neither to the PLF nor the TLF.
The time at which every asymptotic cluster have actually been
emitted ($t_e$) can be calculated within the model by simply tracing back the
dynamical evolution of the reaction. 
See \cite{cher01} and refs. therein for more specific details about the model used and
the cluster time origin analysis of mass asymmetric reactions.

\begin{figure}
\centerline{\epsfig{figure=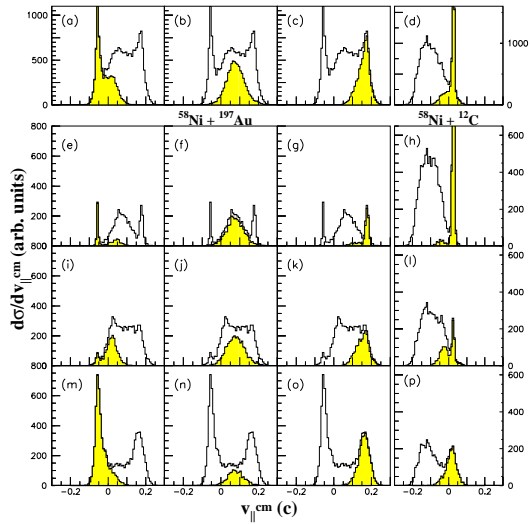,width=7.5cm}}
\caption{ Histograms of $v_{||}^{cm}$ asymptotic distributions. In the first, 
second and third columns, the contributions of the TLF, PEP, and PLF emissions
have been selectively shadowed for the $^{58}$Ni+$^{197}$Au 
reaction ($7$fm $\le b \le 8$fm). The fourth column shows the respective
calculation for the $^{58}$Ni+$^{12}$C reaction ($3$fm $\le b \le 4.5$fm)
shadowing the PLF emissions. Figs. (a)-(d) show total distributions. In Figs. (e)-(h), (i)-(l),
and (m)-(p) the contribution of particles with emission times of
$t_e < 150$ fm/c, $150$ fm/c$ \le t_e < 500$ fm/c, and
$t_e \ge 500$ fm/c after the collision are shown respectively.}
\label{fig_md}
\end{figure}

The extracted emission times for particles originating from the PLF, TLF and PEP components
were used to put into evidence the origins of different regions of the asymptotic velocity space.
Parallel velocity spectra of particles originating from these components were separated
in figure~\ref{fig_md} according to their calculated emission time.
From panels (a)-(c), it can be seen that for the
$^{58}$Ni+$^{197}$Au system, the IV region is populated not only by PEP particles
but also by a supplementary contribution coming mainly from the heavier partner of 
the reaction. Moreover, a different time scale can be associated with each 
kind of contribution. The PEP are mainly light clusters ($Z\le 5$) that were 
configurationally well differentiated from the PLF and TLF very early in
the evolution. On the other hand (see panels (e),(i),(m)), the 
target-like intermediate-velocity clusters were spatially linked to the TLF at PLF-TLF separation time and were
emitted later in the evolution, mainly between  $150$ fm/c and $500$ fm/c.
For the $^{58}$Ni+$^{12}$C case (fourth column of fig.~\ref{fig_md}),
an important contribution to the IV emission coming now from the
PLF can also be recognized. Again, that contribution
is not a prompt one, but takes place in the same mid-range temporal interval
than the one observed in the other reaction. 

These findings suggest two scenarios for the origin of IV particles.
On one hand, the prompt emission at the highly collisional stage of the 
reaction of mainly PEP. On the other, a mechanism where a dynamical
deformation of the heavy partner of the reaction develops,
eventually leading to a scission followed by a 
Coulombian push that projects the
emitted particles towards the intermediate-velocity range, this second
alternative occurring on a longer time scale. 

\begin{figure}
\centerline{
\epsfig{figure=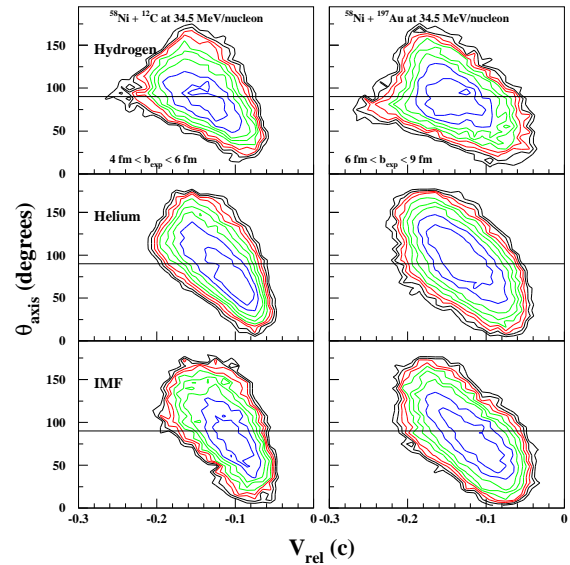,width=7.5cm}}
\caption{\label{fig_thax} $\theta_{axis}$ vs $V_{rel}$ logarithmic contour plots for IV
attributed particles in the $^{58}$Ni+C (left) and $^{58}$Ni+Au (right) systems for
two specific ranges of $b_{exp}$.}
\end{figure}
To investigate these possible origins of IV particles in the data,
a study has been done of the angular relations between their emission direction (evaluated from their
velocity vector in the reconstructed center of mass of all IV particles in the event)
and the reaction axis, defined by the direction of the velocity vector
of the reconstructed QP in the system CM reference frame. Figure~\ref{fig_thax}
shows the distribution of the angle between those two vectors
as a function of the relative velocity between the IV particle and the QP residue,
for particles attributed to the IV component. Hydrogen, Helium and IMF ($Z=3-7$) contributions
are shown on the figure.
From collisions of the $^{58}$Ni+Au system, it is possible to
notice firstly the nearly forward-backward symmetry
of the distributions around $\theta_{axis}=90^{o}$ for hydrogens up to IMF's,
as well as a sizeable focusing of $Z=1$ particles at $90^{o}$.
This can be understood as a fast emission of particles and
excited clusters from the overlapping nuclear matter. Excited clusters will decay on
a variable time scale and most of the final products will be released in an almost
field free region, therefore with isotropic angular distribution. However, the prompt
ejected particles following early nucleon-nucleon collisions in the overlap region will be
subject to the Coulomb field of the two heavy nuclei as well as to blocking from
the saturated quantum phase space regions. These particles will therefore be predominantly
emitted on a plane perpendicular to the reaction axis.
In the $^{58}$Ni+C reaction,
forward-backward symmetry around $\theta_{axis}=90^{o}$ is destroyed, as can
be seen from the left panels of figure~\ref{fig_thax}. Along with a narrower distribution,
an enhancement of forward directed IV particles with small relative velocity with the
QP residue is observed for $Z=2$ particles and IMF's.
This could be a signature for the presence of a supplementary process that populates
the IV region from an asymmetric and aligned breakup of the QP.
This production process of IV
particles seems to be less dominant in the $^{58}$Ni+Au reaction since no
such enhancement is observed. These experimental results go
along the molecular dynamics predictions that aligned asymmetric breakup occurs
mainly on the heavier partner side of the IV region \cite{cher01}. The fact that
the effect is not experimentally seen on the QT side of the $^{58}$Ni+Au system is
just a result of inadequate detection thresholds for particles in this region.

\begin{figure}
\centerline{
\epsfig{figure=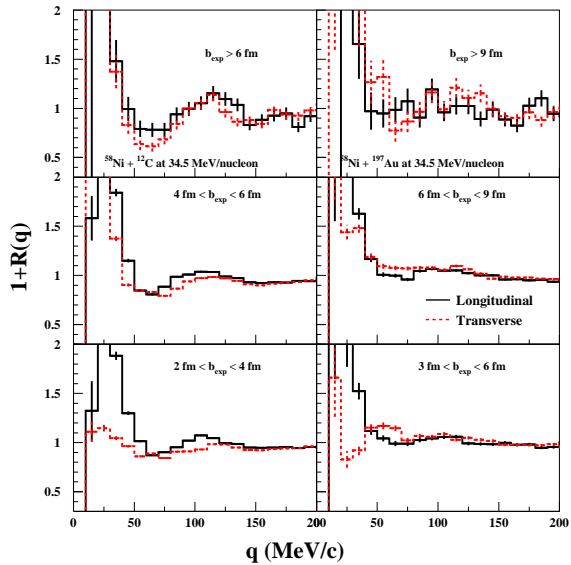,width=7.5cm}}
\caption{\label{fig_corr} $\alpha-\alpha$ correlation functions broken in longitudinal (solid line)
and transverse (dashed line) contributions for IV
attributed particles in the $^{58}$Ni+C (left) and $^{58}$Ni+Au (right) systems.}
\end{figure}
Origins of IV particles have been studied further by two-particle
correlation functions with directional cuts. A similar method was originally used
for proton-proton \cite{lisa93,lisa94} and IMF-IMF \cite{glas94} correlation functions
in order to probe the space-time extent of a compound source in central collisions.
In our case, $\alpha$ particles have been chosen for
this study for their abundance in the IV region and for the reason
that the first unstable excited state of the two-$\alpha$ particles decaying $^{8}$Be nucleus
has a lifetime of about 130 fm/c ($\Gamma$ $\approx$ 1.5 MeV), which is about the time it
takes to the aligned asymmetric breakup process to take place.
The experimental $\alpha-\alpha$ correlation function is defined as
\begin{equation}
1+R(q)=\frac{N_{coinc}(q)}{N_{back}(q)},
\end{equation}
where $N_{coinc}(q)$ is the coincident $\alpha$ pairs yield with relative momentum $q$ and where
$N_{back}(q)$ is the background yield constructed by means of a modified event-mixing technique
for IV attributed particles.
Since final particles in each event have their velocity vectors related to the reaction
axis, a strict application of the standard event-mixing technique introduces unreal contributions
in the background yield for pair of events with different reaction axis directions. Therefore,
by taking advantage of the cylindrical symmetry of the detector apparatus, an improved event-mixing
technique was used which involved a rotation of the second event of a background pair in the plane
perpendicular to the beam, so as to make coincide the $\phi$ angle of the two event reaction axis.
Correlation functions constructed with this technique were measured for two directional cuts in
the emission direction of the two $\alpha$ particles relative to the momentum direction of the
pair in the QP reference frame. Longitudinal correlation functions stand for angles between
the two directions ranging from $0^{o}$ up to $60^{o}$, whereas transverse correlation functions
refer to angles from $60^{o}$ to $90^{o}$. Figure~\ref{fig_corr} shows these correlation
functions for the two studied systems and for different impact parameters.
From the figure, the ground state and the first excited state of the particle
unstable $^{8}$Be nucleus are clearly seen at relative momenta of 25 and 100 MeV/c respectively. The weaker
transverse correlation function for the $^{8}$Be ground state at low $b_{exp}$ was expected from simple Coulomb
trajectory calculations including $^{8}$Be decay and average experimental velocity space relations
for QP and IV particles. It is due to the small
relative angle between the emitted $\alpha$ particles in the laboratory reference frame and to the non-negligeable
size of the detectors leading to double hits that are discriminated out of the correlation function
analysis. However, the calculation showed that multiple hits contribute to a negligeable amount
in the difference between longitudinal and transverse breakup for the first excited state decay
in the IV region.

The approximately equal strengths observed for the longitudinal and
transverse correlation functions in the first excited state region of interest ($80 < q < 120 MeV/c$)
of the $^{58}$Ni+Au reaction
emphasize that $^{8}$Be resonances in the IV region bears no alignement
relations with the QP and are therefore probably isolated from it in space-time at the point
of their breakup. This can be understood as the fast QP leaves the interaction region very early
without dragging nuclear material that can be pushed back in the IV
region later on. However, differences between the strengths of first excited state transverse
and longitudinal correlation functions are observed in the $^{58}$Ni+C reaction
for collisions of experimental impact parameter below $6 fm$.
The Coulomb trajectory calculation of a system, composed initially of a $^{58}$Ni QP emitting
from its surface a $^{8}Be$ nucleus in its first excited state, has shown that tidal forces
were strong enough to flip about $50\%$ of initially emitted longitudinal $\alpha$ pairs in
final transverse ones, whereas only $5\%$ of the reverse case occur.
This effect expected for isotropic $\alpha$ particle decay from $^{8}Be$ in the field of
the remaining QP is however not observed experimentally. On the contrary, longitudinal
decay appears to be dominant in mid-peripheral reactions.
Therefore, this experimentally observed stronger longitudinal correlation function
tends to imply an important initial polarization in the emission of first excited state $^{8}Be$ resonances in the
longitudinal direction.
This can possibly occur from a deformation of the QP in the reaction axis direction, thus constraining
alignement conditions for the $^{8}Be$ resonance formation. In mid-peripheral reactions, these more
important deformations will lead the QP to a breakup configuration that will ultimately push back
some of the material in the IV region.

This deformation breakup effect populating the IV region with light
fragment and particles is associated to the delayed aligned asymmetric breakup observed
in the molecular dynamics simulation \cite{cher01} and in the experimental $\theta_{axis}$ distributions
for the $^{58}$Ni+C system. More important deformations are therefore expected on the heavier
partner side in a mass asymmetric heavy ion reaction. This can be a result of the different
surface boundary conditions at the two poles of the reseparating system. An asymmetrically
shaped neck will breakup earlier on the lighter partner side and will stay attached to the
heavier partner for a longer time, potentially leading it to breakup in a delayed aligned
asymmetric fashion.

In conclusion, we have demonstrated in this paper the existence of two mainly different origins for
intermediate-velocity particle and fragment production. The use of a source reconstruction
technique has permitted to study on an event-by-event basis the characteristic angular emission
pattern of the IV particles as well as their relation with the remaining QP in two-particle correlations.
Important differences between two entrance channel mass asymmetric reactions were observed on the QP side of the
intermediate-velocity space.
With help of time-based cluster
recognition algorithm applied to molecular dynamics simulation \cite{cher01}, it has been possible
to determine the time scales associated to two different phenomena. The first origin
is related to prompt nucleon-nucleon collisions that occur in the overlap zone of the two
colliding nuclei. These processes will eject light particles and excited clusters out of
the overlap on a very short time scale of the order of the reseparation time. Excited clusters
ejected at this stage will however emit particles on a longer time scale. The second origin
of IV particle production is related to the collective
motion of nucleons at the perturbed ends of the QP and QT. Important deformations will be
carried by the heavier partner of the collision and will lead it to a mass asymmetric
breakup aligned along the reseparation axis. This is expected to happen after a delay
of the order of 150-500 fm/c. In entrance channel mass symmetric collisions, molecular
dynamics calculations predict this second origin of IV particles to
happen on both QP and QT. Recent results for the heavier system $^{116}$Sn+$^{93}$Nb
at 29.5 MeV/nucleon presenting the necessity of adding a surface emission component to the neck
component at mid-velocity are compatible with this expectation \cite{pian01}.
It can also be noted that the phenomenon of delayed aligned asymmetric breakup is
potentially related to previously observed processes of dynamical projectile splitting \cite{olmi80}
and fast asymmetric fission \cite{casi93} at lower bombarding energies.
It will be interesting in future analysis and experiments to
study the isospin dependance of the IV products according to their
specific origin, in order to disentangle shape, thermal and chemical equilibration
processes.
In fact, a possible chemical motion in reaction to the slow surface/volume deformations
involved in the delayed aligned asymmetric breakup mechanism could explain in part the
observation of higher $N/Z$ ratio of mid-rapidity IMF's relative to total systems with
no initial neutron enrichment \cite{laro00}. 

\section*{Acknowlegments}
This work has been supported in part by the NSERC, FCAR, University of
Buenos Aires and CONICET.

\end{document}